\documentclass[preprint,8pt]{elsarticle}

\usepackage{amssymb}
\usepackage{tabularx}

\usepackage{lineno}

\journal{Arxiv}

\begin{document}
\begin{frontmatter}

\title{Dosimetric comparison of the BNCT treatment planning performances when using a nnU-NET to automatically segment Glioblastoma Multiforme}



\affiliation[inst1]{organization={Università diPavia},
            addressline={via A. Bassi 6}, 
            city={Pavia},
            postcode={27100}, 
            state={Italy},
            }

\author[inst1]{Cristina Pezzi}
\author[inst1,inst2]{Francesco Morosato}
\author[inst1,inst2,inst3]{Barbara Marcaccio}
\author[inst1,inst2]{Silva Bortolussi}
\author[inst2]{Ricardo Luis Ramos}
\author[inst2]{Valerio Vercesi}
\author[inst2]{Ian Postuma}
\author[inst2]{Setareh Fatemi}

\affiliation[inst2]{organization={Istituto Nazionale di Fisica Nucleare, Unit of Pavia},
            addressline={via A. Bassi 6}, 
            city={Pavia},
            postcode={27100}, 
            state={Italy},
            }
\affiliation[inst3]{organization={Universidad Nacional de San Martin},
addressline={ 25 de Mayo y Francia}, 
         city={Buenos Aires},
         postcode={1650}, 
            state={Argentina}
           }            
\begin{abstract}
This work presents a preliminary evaluation of the use of the convolutional neural network nnU-NET to automatically contour the volume of Glioblastoma Multiforme in medical images of patients. The goal is to assist the preparation of the Treatment Planning of patients who undergo Boron Neutron Capture Therapy (BNCT). BNCT is a binary form of radiotherapy based on the selective loading of a suitable 10-boron concentration into the tumour and on subsequent low-energy neutron irradiation. The selectivity of the therapeutic effect is based on the capacity of the boron drug to target preferentially cancer cells, thus triggering the neutron capture only in the tumour and depositing there a lethal dose. Even if the tailoring of the beam to the tumour volume is less crucial for BNCT than for other radiation therapies, a proper delimitation of the tumour volume is needed to assess a safe and effective dosimetry. In clinical application the contour must be manually decided by the physician, however, a tool to automatically define important structures such as the Gross Tumour Volume (GTV) and the Organs At Risk (OAR) would be beneficial to enable medical physicists assessing preliminary positioning and simulated dosimetry before the approval or possible changes introduced by the radiotherapist. Moreover, an initial contouring may speed up the work of the physician. 
The nnU-NET was trained and tested and its performance was evaluated through different parameters such as the Dice Coefficient. To assess a more meaningful evaluation for BNCT, for the first time, this work analyzed the difference of the clinical dosimetry in 16 patients using the manual and the automatic contoured images. Results show that the neural network performs well in assisting BNCT dose calculation, predicting minimum, maximum and mean dose compatible with those obtained in the manual-obtained contours. The work also highlights some limitations, especially due to the relative low number of clinical cases used for the training in this study. Discussion poses the bases for the next work necessary to fully exploit the potential of such a strategy.  

\end{abstract}



\begin{keyword}
Deep Learning \sep BNCT \sep TPS \sep isoeffective dose
\end{keyword}

\end{frontmatter}


\section{Introduction}
\label{sec:Introduction}
\begin{figure}[h!]
    \centering
    \includegraphics[width=0.98\linewidth]{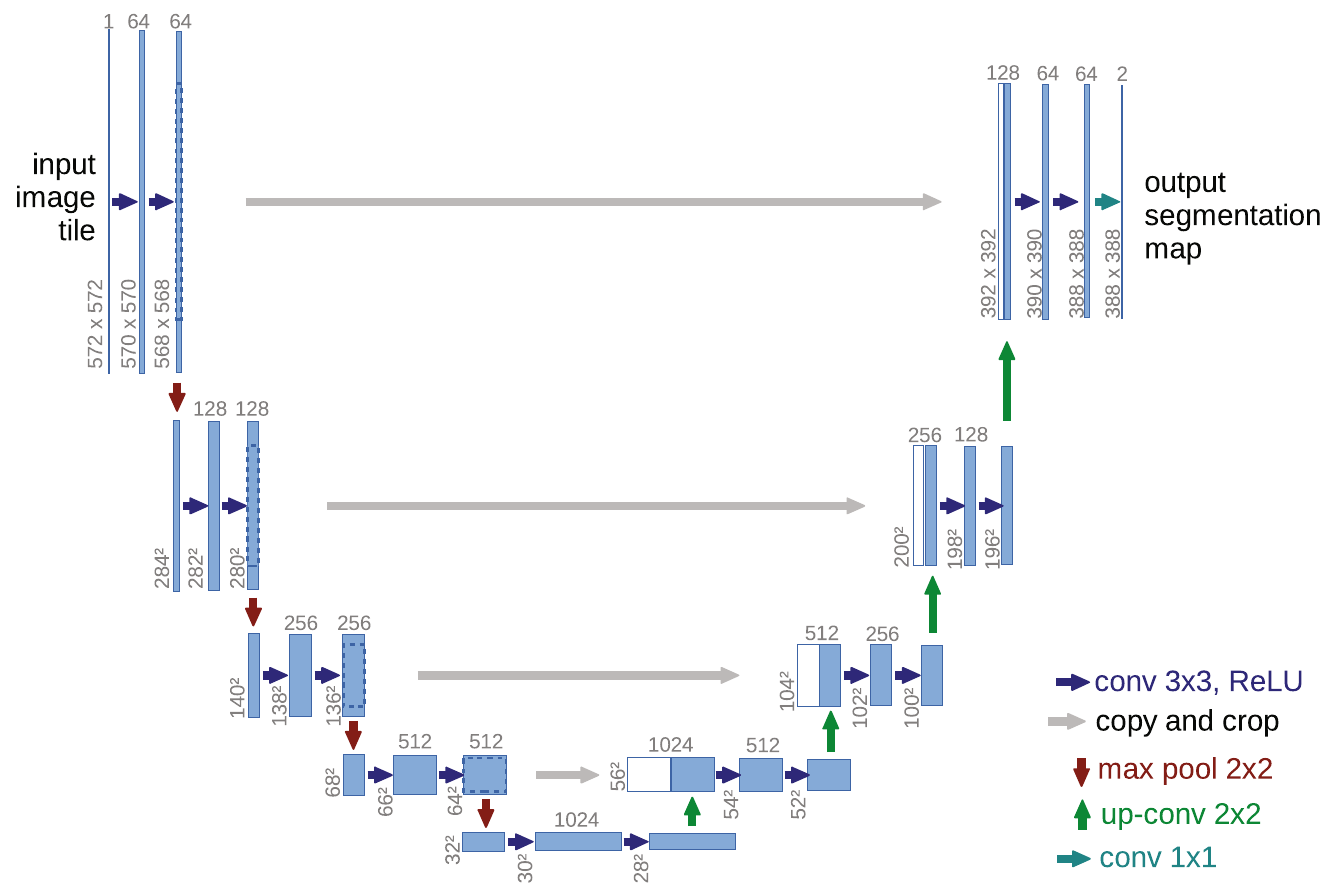}
    \caption{U-NET architecture (example for $32\times32$ pixels in the lowest resolution). Each blue box corresponds to a multi-channel image. The number of channels is denoted on top of the box. The x-y-size is provided at the lower left edge of the box. White boxes represent copied feature maps, that are concatenated to the feature maps represented by the blue box. The arrows denote the different operations. Taken from~\cite{Ronne1}.}
    \label{fig:Unet}
\end{figure}

Boron Neutron Capture Therapy (BNCT) is a binary therapy based on the thermal neutron capture reaction by Boron-10 ($^{10}B$). The high capture cross section of $^{10}B$ for thermal neutrons renders the therapy feasible when the element is loaded in the tumour target with a high ratio respect to the normal tissue. The products of this capture reaction are an alpha particle and a $^{7}Li$-nucleus, both have high Linear Energy Transfer and a combined range in soft tissue comparable to the cell dimension. Therefore, all the energy carried by these two particles is deposited in the target cell ensuring high efficacy to the therapy \cite{IAEA-TecDoc2023}.\\
In recent years many accelerator-based BNCT (ab-BNCT) clinical facilities are operative in various countries of the world, e.g. China, South Korea, Japan and Finland \cite{China, Korea, Matsumura2023, Porra2023} and others are under development in many other countries including Italy\cite{Anthem}. The most commonly treated tumours are Glioblastoma Multiforme (GBM) and Head and Neck Tumours (H\&N). Treatment of H\&N tumours is now part of the National Japanese Health System \cite{Matsumura2023}.\\
Currently Glioblastoma Multiforme is the most common malignant primary tumour of the brain and is one of the most aggressive, the median survival for GBM patients is around 15 months \cite{cancers14102412}. BNCT has been proposed as a treatment method for GBM and numerous studies have been carried out \cite{kawabata2021accelerator,miyatake2020boron}.\\
In this framework this study focuses on GBM cases to apply Artificial Intelligence Methods to accurately and automatically segment the Gross Tumour Volume (GTV).\\
Artificial Intelligence methods and in particular Deep Learning Models (DL) have been successfully applied to image segmentation, in particular U-NET models have become the gold standard to automatically segment medical images \cite{azad2022medical}. The U-NET shown in Fig.\ref{fig:Unet} is a Fully Convolutional Network (FCN), because it does not have any dense layer, but only convolutional ones. It is composed by two parts, the Contracting path or Encoder on the left and the Expansive path or Decoder on the right. 
Every path is the repetition of the same architecture module repeated with different characteristics. 

The innovation of the U-NET consists in how it retrieves the lost spatial information of the input image, that is the expanding path. In this study, the U-NET used is the nnU-NET \cite{isensee2021nnu}, a DL method for segmentation that autonomously configures itself on the whole segmentation task pipeline for any new datasets. This includes decisions about data pre-processing, NN architecture, training and data post-processing for any task. This model is the state of the art in the segmentation field and it has shown better performances than other specialized NNs during the international biomedical segmentation competitions~\cite{isensee2021nnu}.\\
The automatically segmented GTV's using the nnU-NET have been compared to the segmentation performed by a radiologist. However, to obtain relevant information for the BNCT community, this study had the aim to evaluate the performances and the limitations of using automatically segmented images from the dosimetric point of view.\\

\section{Materials and Methods}
The images used in this work have been obtained from a public database: The Cancer Imaging Archive (TCIA) \cite{TCIA}. From all datasets that can be found on TCIA related to Glioblastoma Multiforme cases, only one was chosen for this study.
The GLIS-RT dataset \cite{shusharina2021glioma} is the only GBM datsaset provided with CT images and RT\_STRUCT files, which are sets of anatomical structures identified on radiological images for the planning and delivery of radiotherapy. This collection consists of 230 cases of GBM and low-grade Glioma patients treated with surgery and adjuvant radiotherapy at Massachusetts General Hospital. All images are provided with the radiotherapy targets, the GTV and the clinical target volume (CTV) manually delineated by the radiation oncologist. Of the 230 cases displayed, 198 are GBM patients and these were used for the training and the testing of the nnU-NET.\\
To be suitable for the training and testing of the neural network the DICOM images of the dataset were converted in the NIFTI format. Moreover, to focus on the tumour and reduce computational time, a boundary box algorithm was used to reduce the image dimensions to include only the head of the patient.\\
The images were then used for the nnU-NET, the network architecture was configured to suit the characteristics of CT imaging, while the training and the validation stages were closely monitored adjusting weights and hyperparameters as necessary. The 198 cases have been automatically sorted by the network, allocating 152 (76.7\%) for the training and validation processes and the remaining 46 (23.3\%) for the testing phase.\\
The evaluation of the performances of the network involves the following metrics and indexes to assess the segmentation accuracy and to compare the given ground-truth volume (G) and the NN predicted volume (S).
\begin{itemize}
    \item[(i)] \textbf{\textit{Dice Coefficient (DC)}} which measures the similarity between two volumes. It ranges between 0 and 1, where 0 is obtained when there is no similarity, and 1 means perfect superposition.
\[  DC = \frac{2 \cdot (G \cap S)}{G+S}   \]
    \item[(ii)] \textbf{\textit{Geometrical Miss Index (GMI)}}, which corresponds to the fraction of the ground-truth volume G that is not predicted. Also GMI ranges between 0 and 1, where a higher GMI indicates a larger geometric discrepancy between the real and the predicted volumes. 
\[  GMI = \frac{G - (G \cap S)}{G}   \]    
    \item[(iii)] \textbf{\textit{Discordance Index (DI)}}, which  is the fraction of the predicted volume S that does not belong to ground-truth volume G. It measures the over-contouring of the NN respect to the manual segmented image. 
\[  DI = \frac{S - (G \cap S)}{S}   \]
\end{itemize}
Of the 46 cases used for the testing of the algorithm, 16 were selected for this analysis. This subset comprised cases with both low and high Dice Index and with variable tumour sizes as shown in Tab.\ref{tab:17_pazienti}.\\

\begin{table}[h!]
\begin{tabular}{|c |c |c |c |c |c |}
 \hline
    Patient & \textbf{DC} & \textbf{GMI} & \textbf{DI} & \begin{tabular}[c]{@{}c@{}}\textbf{Volume true}\\ ($cm^3$)\end{tabular}  & \begin{tabular}[c]{@{}c@{}}\textbf{Volume NN} \\ ($cm^3$)\end{tabular}  \\ \hline               
\textbf{GBM\_0071}   & 0.296          & 0.002                           & 0.774                      & 33.681                & 32.832                   \\ \hline
\textbf{GBM\_0057}   & 0.369          & 0.038                           & 0.734                      & 29.267                & 128.957              \\ \hline
\textbf{GBM\_0137}   & 0.417          & 0.044                           & 0.725                      & 31.274                & 113.013               \\   \hline
\textbf{GBM\_0202}   & 0.427          & 0.105                           & 0.574                      & 11.446                & 39.766               \\   \hline
\textbf{GBM\_0096}   & 0.577          & 0.391                           & 0.078                      & 124.71                & 261.85                \\   \hline
\textbf{GBM\_0217}   & 0.688          & 0.097                           & 0.333                      & 253.558               & 152.465             \\    \hline
\textbf{GBM\_0001}   & 0.767          & 0.078                           & 0.306                      & 75.297                & 101.889             \\  \hline
\textbf{GBM\_0020}   & 0.777          & 0.039                           & 0.348                      & 30.747                & 42.127              \\   \hline
\textbf{GBM\_0157}   & 0.777          & 0.176                           & 0.237                      & 107.539               & 158.435     \\ \hline
\textbf{GBM\_0192}   & 0.793          & 0.231                           & 0.150                      & 34.354                & 37.096                  \\ \hline
\textbf{GBM\_0225}   & 0.807          & 0.199                           & 0.152                      & 202.057               & 182.820               \\ \hline
\textbf{GBM\_0022}   & 0.824          & 0.137                           & 0.177                      & 61.243                & 57.871                \\  \hline
\textbf{GBM\_0129}   & 0.843          & 0.073                           & 0.214                      & 134.324               & 140.839                  \\  \hline
\textbf{GBM\_0144}   & 0.851          & 0.093                           & 0.173                      &  89.635                & 105.703                   \\  \hline
\textbf{GBM\_0060}   & 0.853          & 0.076                           & 0.100                      & 104.607               & 114.451               \\   \hline
\textbf{GBM\_0128}   & 0.912          & 0.042                           & 0.842                      & 207.673               & 213.312               \\  \hline

\end{tabular}
\caption{\textit{Volumetric dimensions and values of DC, GMI, and DI for each analyzed patient. Volumes are referenced for both manually segmented GTVs and those artificially generated by the the neural network.}}
\label{tab:17_pazienti}
\end{table}

\begin{figure}[!ht]
    \centering
    \includegraphics[width=0.8\textwidth]{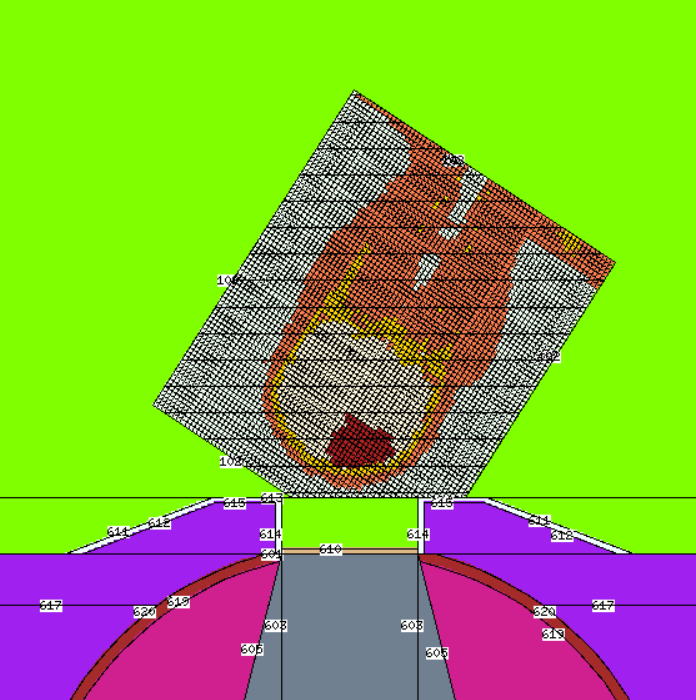}
    \caption{\textit{Simulation of the treatment planning of one of the cases analyzed in this work. Green:air, purple, pink, grey:materials of the Beam Shaping Assemply, as described in \cite{postuma2021novel}.}} 
    \label{fig:positioning}
\end{figure}

To evaluate the performances of the neural network segmentation, in addition to the  indexes listed above, a dosimetric comparison was performed to test the suitability of automatic contouring in BNCT. Dose distributions and values were calculated in the segmentations made by the radiation oncologist and in the masks obtained from the neural network.\\
To simulate the Treatment Planning we used the IT\_STARTS Treatment Planning System (TPS) developed by INFN (publication in preparation).
For the simulation of a BNCT treatment, it is necessary to position the patient model inside the irradiation room, in the most advantageous orientation with respect to the neutron beam port.\\
For the irradiation simulation, the neutron source designed for the Radiofrequency Quadrupole accelerator (RFQ) designed and built in Italy by INFN, delivering a 5 MeV, 30 mA proton beam~\cite{acceleratore}. A facility consisting of the RFQ, a beryllium target for neutron production and a beam shaping assembly for moderation and collimation \cite{postuma2021novel}, will be built in Caserta (Italy) in the frame of the PNC-PNRR ANTHEM project.
 In order to achieve an optimal dose distribution, several strategies can be applied, and medical considerations must be taken into account regarding the protection of sensitive structures. In this work, 
 we minimised the distance of the centre of mass of the GTV segmented by the radiation oncologist from the beam-port. This corresponds to maximising the irradiation of the GTV, regardless possible limitations of the surrounding structures. This choice is justified because the goal of this study is not the evaluation of BNCT effectiveness and therapeutic potential in specific cases, but rather to evaluate the relative differences of treatment plans, once fixed the irradiation position. Figure \ref{fig:positioning} shows the geometry of the beam port and the voxelized patient model oriented to comply with the criterion described above.

 For photon-equivalent dose calculation, IT\_STARTS allows the input of radiobiological data for both the RBE-weighted dose model and the photon isoeffective dose model \cite{isoeff}.
The absorbed dose in the OAR is calculated applying the RBE-weighted dose model and assuming the RBE (Relative Biological Effectiveness) and CBE (Compound Biological Effectiveness) parameters given by \textit{Coderre}~\cite{coderre_morris}, following the protocol applied in clinics, i.e. $CBE = 1.35$, $RBE_{th} = RBE_{f} = 3.2$, $RBE_{\gamma}=1$. The dose in the GTV is calculated using the isoeffective dose model that was proven to better express the dose-effect relation in BNCT, expressing the mixed-field dose in photon equivalent units with higher predictive power \cite{gonzalez2017PMB}. The radiobiological parameters to implement the model were taken from dose-survival studies using the human Glioblastoma U87 cell line. Monolayer cell cultures were irradiated with photons, neutrons and neutrons in presence of boron, at the TRIGA Mark II reactor of the Pavia University (data to be published).\\
The calculation of the treatment based on the radiation oncologists' segmentation was then compared to the one obtained using the mask generated by the nnU-NET. The simulations produced the Dose Volume Histograms in the GTVs and it was thus possible to extract the minimum absorbed dose (the dose absorbed by 98\% of the GTV volume), the mean dose and the maximum absorbed dose (the dose absorbed by 2\% of the volume).\\

\section{Results}
\label{sec:Results}
\begin{figure}[!ht]
    \centering
    \includegraphics[width=0.95\textwidth]{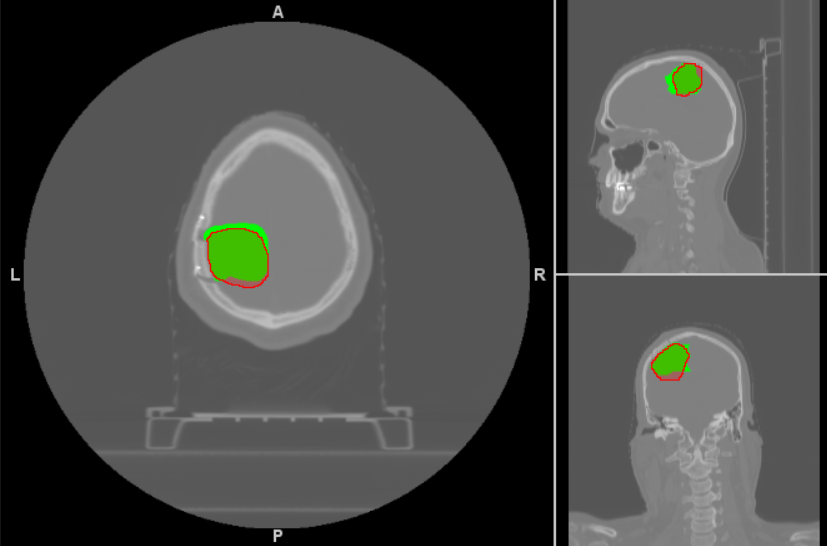}
    \caption{\textit{CT imaging of a representative case in which the manually segmented GTV (green) and the contouring performed by the neural network (red) are superposed.}} 
    \label{fig:example}
\end{figure}

The Treatment Planning simulation was conducted on the 16 available cases of GBM patients. Table \ref{tab:big_tabella} lists  the irradiation time obtained in the two segmentations, fixing the patient position by using as a reference the tumour segmented by the radiation oncologists. Moreover, the value of minimum, mean and maximum dose to the GTV are reported in both cases. 
Figure \ref{fig:example} shows a graphical representation of a patient imaging overlaid with the manually segmented GTV (green) and
the one segmented by the neural network (red). In this case, the Dice Coefficient was high, proving a good superposition of the manual and the automatic contouring.

\begin{figure}[!ht]
    \centering
    \includegraphics[width=0.95\textwidth]{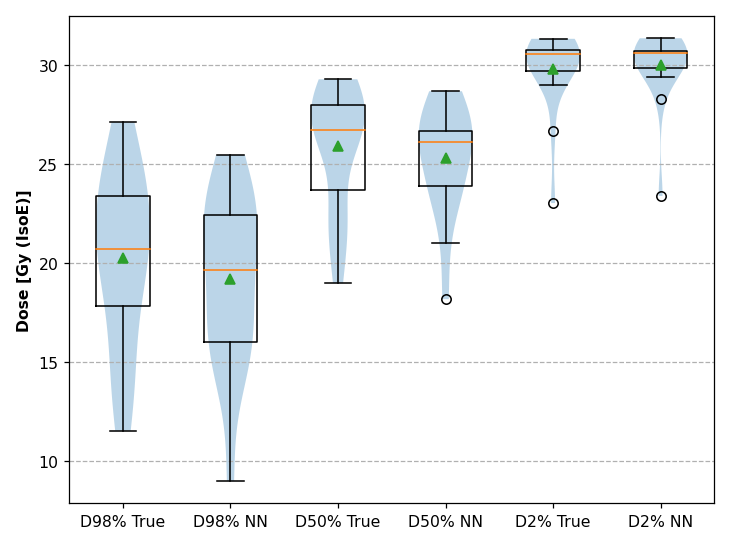}
    \caption{\textit{Box-violin plot of the distribution of dose values in the GVTs. From left to right: values of minimum, mean, and maximum doses deposited in the GTVs. The mean value is the green triangle and the median is graphically represented by the orange line.}} 
    \label{fig:dosii}
\end{figure}

Figure \ref{fig:dosii} shows the box-violin plots of the minimum, mean and maximum dose for both manual and automatic segmentation. These plots show the mean value and the median for each parameter taking into account the whole dataset.\\

\begin{table}[h!]
\begin{tabular}{|c|cc|cccccc |}
\hline
\textbf{Patient}                & \begin{tabular}[c]{@{}c@{}}\textbf{t}\textsubscript{irr}\\ \textbf{(min)}\end{tabular} & \begin{tabular}[c]{@{}c@{}}\textbf{t}\textsubscript{irr} \\ \textbf{(min)}\end{tabular} & \multicolumn{1}{c}{\begin{tabular}[c]{@{}c@{}}\textbf{D98\%} \\ \textbf{(Gy)}\end{tabular}} & \multicolumn{1}{c}{\begin{tabular}[c]{@{}c@{}}\textbf{D98\%}\\ \textbf{(Gy)}\end{tabular}} & \multicolumn{1}{c}{\begin{tabular}[c]{@{}c@{}}\textbf{D50\%} \\ \textbf{(Gy)}\end{tabular}} & \multicolumn{1}{c}{\begin{tabular}[c]{@{}c@{}}\textbf{D50\%}\\ \textbf{(Gy)}\end{tabular}} & \multicolumn{1}{c}{\begin{tabular}[c]{@{}c@{}}\textbf{D2\%} \\ \textbf{(Gy)}\end{tabular}} & \multicolumn{1}{c|}{\begin{tabular}[c]{@{}c@{}}\textbf{D2\%} \\ \textbf{(Gy)}\end{tabular}} \\
                                          & \textbf{True}                                                                                                            & \textbf{NN}                                                                                                               & \multicolumn{1}{c}{\textbf{True}}                                                  & \multicolumn{1}{c}{\textbf{NN}}                                                   & \multicolumn{1}{c}{\textbf{True}}                                                  & \multicolumn{1}{c}{\textbf{NN}}                                                   & \multicolumn{1}{c}{\textbf{True}}                                                 & \multicolumn{1}{c|}{\textbf{NN}}                                                                                                                  \\ \hline \hline

\begin{tabular}[c]{@{}c@{}}\textbf{GBM}\\ \textbf{0071}\end{tabular}
 & 29.60                                                                                                           & 29.58                                                                                                            & 18.3                                                                      & 18.6                                                                     & 20.7                                                                      & 21.0                                                                     & 23.0                                                                     & 23.4                                                                                                                                   \\ \hline
\begin{tabular}[c]{@{}c@{}}\textbf{GBM}\\ \textbf{0057}\end{tabular}      & 32.06                                                                                                           & 34.77                                                                                                            & 21.7                                                                      & 20.9                                                                     & 26.6                                                                      & 26.2                                                                     & 29.0                                                                     & 30.2                                                                                                                                   \\ \hline
\begin{tabular}[c]{@{}c@{}}\textbf{GBM}\\ \textbf{0137}\end{tabular}       & 39.20                                                                                                           & 38.30                                                                                                            & 25.4                                                                      & 22.5                                                                     & 29.1                                                                      & 26.7                                                                     & 30.9                                                                     & 30.2                                                                                                                                  \\ \hline
\begin{tabular}[c]{@{}c@{}}\textbf{GBM}\\ \textbf{0202}\end{tabular}       & 30.92                                                                                                           & 31.67                                                                                                            & 27.1                                                                      & 24.0                                                                     & 29.3                                                                      & 28.7                                                                     & 30.5                                                                     & 30.7                                                                                                                                   \\ \hline
\begin{tabular}[c]{@{}c@{}}\textbf{GBM}\\ \textbf{0096}\end{tabular}       & 32.58                                                                                                           & 35.35                                                                                                            & 11.5                                                                      & 9.0                                                                      & 19.0                                                                      & 18.2                                                                     & 26.7                                                                     & 28.3                                                                                                                                   \\ \hline
\begin{tabular}[c]{@{}c@{}}\textbf{GBM}\\ \textbf{0217}\end{tabular}      & 36.98                                                                                                           & 36.35                                                                                                            & 13.3                                                                      & 16.5                                                                     & 21.8                                                                      & 23.5                                                                     & 30.0                                                                     & 29.8                                                                                                                                  \\ \hline
\begin{tabular}[c]{@{}c@{}}\textbf{GBM}\\ \textbf{0001}\end{tabular}       & 36.42                                                                                                           & 35.65                                                                                                            & 20.2                                                                      & 18.8                                                                     & 27.4                                                                      & 26.0                                                                     & 31.1                                                                     & 30.7                                                                                                                                   \\ \hline
\begin{tabular}[c]{@{}c@{}}\textbf{GBM}\\ \textbf{0020}\end{tabular}       & 36.09                                                                                                           & 36.09                                                                                                            & 25.1                                                                      & 23.4                                                                     & 29.0                                                                      & 28.4                                                                     & 31.3                                                                     & 31.3                                                                                                                                   \\ \hline
\begin{tabular}[c]{@{}c@{}}\textbf{GBM}\\ \textbf{0157}\end{tabular}       & 39.93                                                                                                           & 39.81                                                                                                            & 19.9                                                                      & 17.6                                                                     & 26.6                                                                      & 26.0                                                                     & 30.9                                                                     & 30.8                                                                                                                                  \\ \hline
\begin{tabular}[c]{@{}c@{}}\textbf{GBM}\\ \textbf{0192}\end{tabular}      & 30.71                                                                                                           & 30.79                                                                                                            & 24.0                                                                      & 25.5                                                                     & 27.6                                                                      & 27.9                                                                     & 29.4                                                                     & 29.4                                                                                                                                   \\ \hline
\begin{tabular}[c]{@{}c@{}}\textbf{GBM}\\ \textbf{0225}\end{tabular}       & 33.83                                                                                                           & 33.86                                                                                                            & 15.5                                                                      & 15.5                                                                     & 22.8                                                                      & 24.1                                                                     & 30.5                                                                     & 30.7                                                                                                                                   \\ \hline
\begin{tabular}[c]{@{}c@{}}\textbf{GBM}\\ \textbf{0022}\end{tabular}       & 37.30                                                                                                           & 38.05                                                                                                            & 23.3                                                                      & 23.1                                                                     & 28.0                                                                      & 28.4                                                                     & 30.7                                                                     & 31.1                                                                                                                                  \\ \hline
\begin{tabular}[c]{@{}c@{}}\textbf{GBM}\\ \textbf{0129}\end{tabular}     & 35.58                                                                                                           & 35.39                                                                                                            & 20.5                                                                      & 20.5                                                                     & 26.4                                                                      & 26.4                                                                     & 30.8                                                                     & 30.7                                                                                                                               \\ \hline
\begin{tabular}[c]{@{}c@{}}\textbf{GBM}\\ \textbf{0144}\end{tabular}       & 33.50                                                                                                           & 33.94                                                                                                            & 22.1                                                                      & 21.9                                                                     & 26.8                                                                      & 26.6                                                                     & 29.7                                                                     & 29.9                                                                                                                                 \\ \hline
\begin{tabular}[c]{@{}c@{}}\textbf{GBM}\\ \textbf{0060}\end{tabular}       & 31.37                                                                                                           & 31.37                                                                                                            & 20.9                                                                      & 22.3                                                                     & 26.6                                                                      & 26.5                                                                     & 29.8                                                                     & 29.7                                                                                                                                  \\ \hline
\begin{tabular}[c]{@{}c@{}}\textbf{GBM}\\ \textbf{0128}\end{tabular}       & 36.96                                                                                                           & 36.96                                                                                                            & 14.4                                                                      & 14.6                                                                     & 22.7                                                                      & 23.0                                                                     & 30.6                                                                     & 30.6                                                                                                                               \\ \hline

\end{tabular}
\caption{\textit{Irradiation time, minimum, mean, and maximum dose values for the 16 analyzed patients. Values were calculated for both manually segmented GTVs (True) and those artificially generated by the the neural network (NN).}}
    \label{tab:big_tabella}
\end{table}

The variation of such parameters was analyzed considering the volume ratio obtained between the medically segmented GTV and the neural network segmented volume.
 To such purpose we also used the Dice coefficient which represents the degree of similarity between the two segmented volumes.\\
A first consideration is that in AI-based GTVs the values of minimum and mean absorbed dose are lower. This tendency to underestimate the minimum and the mean doses seems a direct consequence of the different volume value and a different tumor depth extension predicted by the network. In fact, the deeper the GTV develops, and therefore farther from the beam-port, the lower the minimum absorbed dose is. This difference is less important regarding maximum dose values, which all remain slightly above 30 Gy(IsoE). The maximum absorbed dose, in fact, is located closer to the patient's surface and is more closely related to the prescription, the irradiation time and dose limits, set equally for all patients under examination.

\begin{figure}[!h]
    \centering
    \includegraphics[width=1\textwidth]{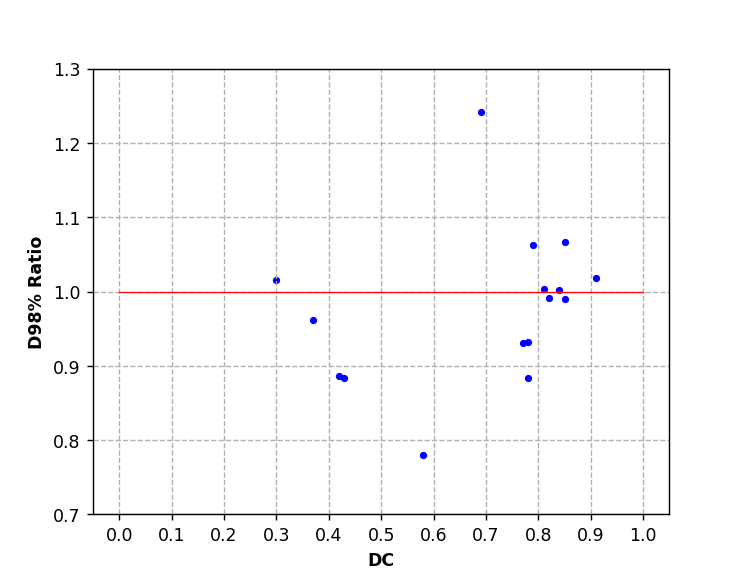}
    \caption{\textit{Plot of the trend of the minimum dose ratio as a function of the Dice Coefficient for the 16 patients analysed.}} 
    \label{fig:d98_vs_dice}
\end{figure}

 In Figure \ref{fig:d98_vs_dice} the minimum dose ratio, which is the value of the minimum dose obtained in the automatic segmentation divided by the minimum dose obtained in the manual segmentation, is shown in respect to the Dice Coefficient. The plot shows that the cases where the ratio of minimum doses tends to 1 are mainly situated in the region of high Dice Coefficient and low Geometrical Miss Index and Discordance Index. The reason is strictly related to the definition of these parameters. Higher DC as well as lower GMI and DI are attributable to a good correspondence and overlapping between the two GTVs, it is thus reasonable to assume that the distributions of doses in the two volumes are similar. As expected, a better overlap and a proper volume reconstruction lead to similar minimum doses.
Focusing more on the relationship between the ratio of minimum doses and the Dice Coefficient, two different behaviors are observed in Figure~\ref{fig:d98_vs_dice} depending on the Dice.

\section{Discussion}
\label{sec:Discussion}
These findings support the conclusion that the training of a neural network nnU-NET assuming the Dice Coefficient as the Loss Function can be efficient in achieving adequate segmented medical images. However, it is also observed that, for low Dice values, predicting correct dose distributions becomes more challenging. Clearly, additional factors come into play, such as, for example, the difference in depth-distribution of the GTVs. This is typical of BNCT, where the distribution of the total dose varies significantly as the neutron beam penetrates the patient's body. 
Moreover, some of these complications arise from the limited number of clinical cases used to train the nnU-NET. This work was important to point out that it is challenging for the neural network to generalize the obtained results, to take into account the complexity of the problem, and to achieve accurate outcomes with this number of cases analyzed. However, even with these limitations, this study has demonstrated, for the first time, the possibility to apply Artificial Intelligence algorithms for the BNCT treatment of patients affected by brain tumors. The artificial segmentation can indeed be incorporated into a treatment plan, and a corresponding dose distribution can be efficiently generated.\\
Improvements in GTV contouring with ANN could come from defining alternative loss functions, different from the Dice similarity index. New loss functions may include more evaluation metrics, leading GTV prediction to ponder over multiple coefficients, increasing the ANN learning ability. Metrics that may be added to the loss function are GMI and DI.\\
This study represents an initial approach in applying AI algorithms in the field of BNCT, opening the way to further research in this sector, and to the optimization of the network performance when applied to clinical cases. One of the first considerations is that it is difficult to ascertain whether the differences in minimum, mean and maximum dose are clinically significant. More meaningful figures of merit such as a Tumour Control Probability (TCP) should be calculated to inter-compare the treatment planning results on the basis of its clinical effect. To this end, we are developing a model of TCP for GBM and understanding how to use it to assess the AI segmentation performance.
Moreover, this work assumed the manually contoured structures as ground truth. However, manual contouring itself is subject to uncertainties and inter-observer variability. Contouring performed by a different physician would produce a different segmented GTV, resulting in a different dose distribution. It is thus interesting to compare the variability between different manual contours and the discrepancy observed between ground truth and AI to draw more robust conclusions on the significance of the difference between dose distributions obtained in the manual and in the automatic contouring.\\

Artificial Intelligence may be an important tool for the development of a more efficient and reliable BNCT and its use is being explored for dosimetry and treatment planning \cite{Tang} and in other aspects of the research such as boron concentration evaluation in biological samples \cite{Agus}. However, its potential is still to be explored in this field and a huge amount of scientific work is expected in the future. This contribution represents the starting point to insert an AI module into the software for dose determination in patients.

\section{Conclusions}
\label{sec:Conclusions}
The aim of this work was to assess the suitability of AI algorithms in contouring medical images for BNCT treatment of Glioblastoma Multiforme (GBM). The comparison between manually segmented CT images of GBM for dosimetry purposes and those segmented with the assistance of Artificial Intelligence provides valuable insights into the effectiveness and reliability of automated segmentation techniques in radiotherapy planning.
Automatic contouring will not substitute the work by medical doctors, but it can be a valuable aid in BNCT treatment planning, bringing positive effects both in research and clinical settings. 
In research, more patients can be studied and evaluated, and computational dosimetry can become more refined and robust, towards a better description of the dose distribution in the tissues. New tumour targets can be studied as potential candidates for BNCT from the point of view of dose distribution and therapeutic advantage. In clinics, a more efficient treatment planning procedure would enlarge the number of patients, assisting the physician decision process and alowing for a rapid re-evaluation of dose distributions. In fact, the time-consuming phase of BNCT treatment planning is the positioning and Monte Carlo transport; once the dose values have been obtained they can be re-calculated in new masks, if necessary. This, in turn, would  stimulate new research and a general improvement of the quality of the therapy for enhanced clinical applicability in the future.

\section*{Acknowledgements}
This work was funded by the National Scientific Committee 5 of National Institute of Nuclear Physics (INFN), project AI\_MIGHT.
This work was also supported by the National Plan for NRRP Complementary Investments (PNC, established with the decree-law 6 May 2021, n. 59, converted by law n. 101 of 2021) in the call for the funding of research initiatives for technologies and innovative trajectories in the health and care sectors (Directorial Decree n. 931 of 06-06-2022) - \textbf{project n. PNC0000003 - AdvaNced Technologies for Human-centrEd Medicine (project acronym: ANTHEM)}. This work reflects only the authors’ views and opinions, neither the Ministry for University and Research nor the European Commission can be considered responsible for them.



 \bibliographystyle{elsarticle-num} 
 \bibliography{cas-refs}





\end{document}